\documentclass[twocolumn,a4paper,showpacs,preprintnumbers,amsmath,amssymb,nofootinbib,floatfix]{revtex4}
\input psfig.sty  
\def \be {\begin{equation}} 

\def \ee {\end{equation}} 
\def \bea {\begin{eqnarray}} 
\def \eea {\end{eqnarray}} 

\usepackage{graphicx}
\usepackage{dcolumn}
\usepackage{bm}
\usepackage{epsfig}

\newcommand*{\mysub}[2]{\ensuremath{#1_{\mathrm{#2}}}}

\newcommand*{\Mgas}{\mysub{M}{gas}}

\newcommand*{\ltsim}{\ {\raise-.75ex\hbox{$\buildrel<\over\sim$}}\ }
\newcommand*{\gtsim}{\ {\raise-.75ex\hbox{$\buildrel>\over\sim$}}\ }
\newcommand*{\proptosim}{\ {\raise-.75ex\hbox{$\buildrel\propto\over\sim$}}\ }

\begin{document}
\title{Limits on evolution of the fine-structure constant in runaway dilaton  models from  Sunyaev-Zeldovich Observations}
\author{R. F. L. Holanda$^{1,2,3}$ } \email{holanda@uepb.edu.br}
\author{L. R. Cola\c{c}o$^2$} \email{colacolrc@gmail.com}
\author{R. S. Gon\c{c}alves$^{4}$}\email{rsousa@on.br}
\author{J. S. Alcaniz$^4$}\email{alcaniz@on.br}
\affiliation{\\$^1$Departamento de F\'{\i}sica, Universidade Estadual da Para\'{\i}ba, 58429-500, Campina Grande - PB, Brasil\\
$^2$ Departamento de F\'{\i}sica, Universidade Federal do Rio Grande do Norte, 59300-000, Natal - RN, Brasil ,\\
$^3$ Departamento de F\'{\i}sica, Universidade Federal de Campina Grande, 58429-900, Campina Grande - PB, Brasil,\\
$^4$ Observat\'orio Nacional, 20921-400, Rio de Janeiro - RJ, Brasil}

\date{\today}

\begin{abstract}

{  In this paper, new bounds on possible variations of the fine structure constant, $\alpha$, for a class of runaway dilaton  models are performed. By considering a possible evolution with redshift, $z$, such as $\frac{\Delta\alpha}{\alpha}=-\gamma\ln(1+z)$, where in $\gamma$ are the physical properties of the model, we constrain this parameter by using a deformed cosmic distance duality relation jointly with gas mass fraction (GMF) measurements of galaxy clusters and luminosity distances of type Ia supernovae. The GMF's used in our analyses are from cluster mass data from 82 galaxy clusters in the redshift range $0.12<z<1.36$, detected via the Sunyaev-Zeldovich effect at 148 GHz  by the Atacama Cosmology Telescope. The type Ia supernovae are from the Union2.1 compilation.  We also explore the dependence of the results from four models used to describe the galaxy clusters. As a result no evidence of variation was obtained.}

\end{abstract}
\maketitle

\section{Introduction}

The Sunyaev-Zeldovich effect (SZE) is a secondary anisotropy into the cosmic microwave background radiation (CMB) temperature\cite{sze}. It is produced by the inverse Compton scattering of the CMB photons passing through a population of hot electrons in galaxy clusters. This effect encode information about the distribution of dark matter and gas throughout the Universe, being especially important at high redshifts ($z > 1$) where the cosmological model and abundance of clusters are critically correlated. A remarkable feature of the distortion is a decrement in low frequency ($<218$ GHz) and an increment in higher frequency ($> 218$ GHz) in the CMB intensity (see Refs.\cite{bir,car} for excellent reviews).

In the past decade, some authors obtained results from the SZE science by using the Owens Valley Radio Observatory and the Berkeley-Illinois-Maryland Association interferometric arrays. For instance, the Refs.\cite{ree,bon,def} considered a jointly analysis with X-ray surface brightness of galaxy clusters and estimated the angular diameter distance (SZE/X-ray technique) of galaxy clusters by using different assumptions to their morphology. In Ref.\cite{lar} the measurements of gas mass fraction  as well as the scaling relations via SZE were explored. Cosmological parameters also were inferred by using the SZE and other cosmological data in Refs.\cite{cun,hol}. However, since the signal intensity of the SZE is very thin, $10^{-5}$, its potential as a cosmological tool has been explored only in recent years. 
Currently, the South Pole Telescope \cite{haa,ble,sem}, the Atacama Cosmology Telescope \cite{has,mar} and the Planck satellite \cite{ade,ade2,agh} have detected through the SZE about 1000 galaxy clusters including more than hundreds of new galaxy clusters previously unknown by any other observational technique and put tighter constraints  on cosmological parameters. 

The SZE observations also allow us to test the adiabatic evolution of the temperature of the cosmic microwave background (CMB), a key prediction of standard cosmology. Actually, the SZE is redshift independent only if there is no injection of photons into CMB, i.e., if its temperature evolution law is given by $T_{CMB}(z)=T_0(1+z)$. By taking a more general relation, such as $T_{CMB}(z)=T_0(1+z)^{1+\delta}$,  recent analyses tested the  evolution of the CMB temperature through different techniques and confirmed the standard relation, $\delta \approx 0$ (see Refs.\cite{bat,lam,suz}).{  However, in Ref.\cite{jens}, the author showed that  the SZE have limited applicability in these kind of tests.}

On the other hand, it has been showed that the combination of X-ray surface brightness of galaxy clusters with their SZE measurements also can be used to investigate fundamental physics as well as testing results from standard cosmology. For instance, the Ref.\cite{uza} showed that the so-known technique SZE/X-ray of measuring angular diameter distance (ADD) of galaxy clusters depends on the cosmic distance duality relation (CDDR), $D_L(1+z)^{-2}/D_A=\eta=1$, where $D_L$ and $D_A$ are the luminosity and angular diameter distances, respectively. This relation is a fundamental one from cosmology \cite{ete,ell}, requiring only that source and observer are connected by null geodesics in a Riemannian spacetime and that the number of photons is conserved. This relation has been verify at least within 2$\sigma$ c.l. (see table I in Ref.\cite{hol3}, other studies testing the CDDR can be found in Ref.\cite{out}). 

More recently, the authors from Refs.\cite{hol4,hol5} showed that SZE and X-ray observations also can be combined to investigate possible variations of the fundamental constants, specifically, the fine structure constant, $\alpha = e^2/c \hbar$, where $e$ is the charge of electron, $\hbar$ is the Planck constant and $c$ is the speed of the light. Constraints on variations of $\alpha$ for a class of dilaton runaway models were discussed. In these models of $\alpha$ variation,  the relevant parameter is the coupling of the dilaton field to hadronic matter. Several observational analyses  have been performed to study  possible variations of $\alpha$ and to establish bounds on such variations, namely: astronomical observations, based on mainly on the analysis of high-redshift quasar absorption systems\cite{web}; and local methods, based on atomic clocks with different atomic numbers\cite{fis,pei,ros}. An interesting recent debate was done by Refs.\cite{kin,whi} on a possible $\alpha$ variation by using Keck/HIRES and VLT/UVES observations. No important deviation was verified with these observations.

It is important to stress that the X-ray surface brightness depends on the CDDR  and the $\alpha$ while the SZE depends exclusively on  $\alpha$. In this way, a theoretical {  result from Ref.\cite{hee}, $\eta^2(z)=\varphi(z)$ (where $\varphi(z)-1=\Delta \alpha/\alpha$),} was used in order to put limits on $\varphi(z)$. Indeed, the authors of Ref.\cite{hee} showed that for a large class of theories arising from modifications of gravity via the presence of a scalar field with a multiplicative coupling to the electromagnetic Lagrangian, violations of CDDR, of CMB temperature law and variations of $\alpha$ are  intimately and unequivocally linked.

In this paper, we obtained constraints on variations of $\alpha$  for a class of dilaton runaway models  by using  galaxy cluster masses  from  the Atacama Cosmology Telescope (ACT) data  obtained via their ESZ observations and type Ia Supernovae from Union2.1 compilation \cite{suzu}. More precisely, we use measurements of gas mass fraction, $f_{gas}$, obtained from 82 points of galaxy cluster mass \cite{has}. The $f_{gas}$ estimated for each cluster in the sample was calculated by using a semi-empirical relation presented by Ref.\cite{cha}, where the observed gas fraction in galaxy clusters with $z<0.09$ was verified to be  a function of the total mass, $M$. The masses of clusters are those corresponding to  $M_{500}$, defined as the mass measured within the radius $R_{500}$. Since these measurements  depend on the physical model  of the intracluster gas, the  ACT team adopted four models (see Sec. III for details). So, as an extra bonus, we also  verify the dependence of our results  with  the methods used to infer $M_{500}$.

The paper is organized as follows. In Section II,  we briefly describe the samples used in our analyses. In Section III  we describe our method. In section IV, we perform the analyses. Finally, the discussions and conclusions are given in Section V.

\begin{figure*}
\centering
\includegraphics[width=0.47\textwidth]{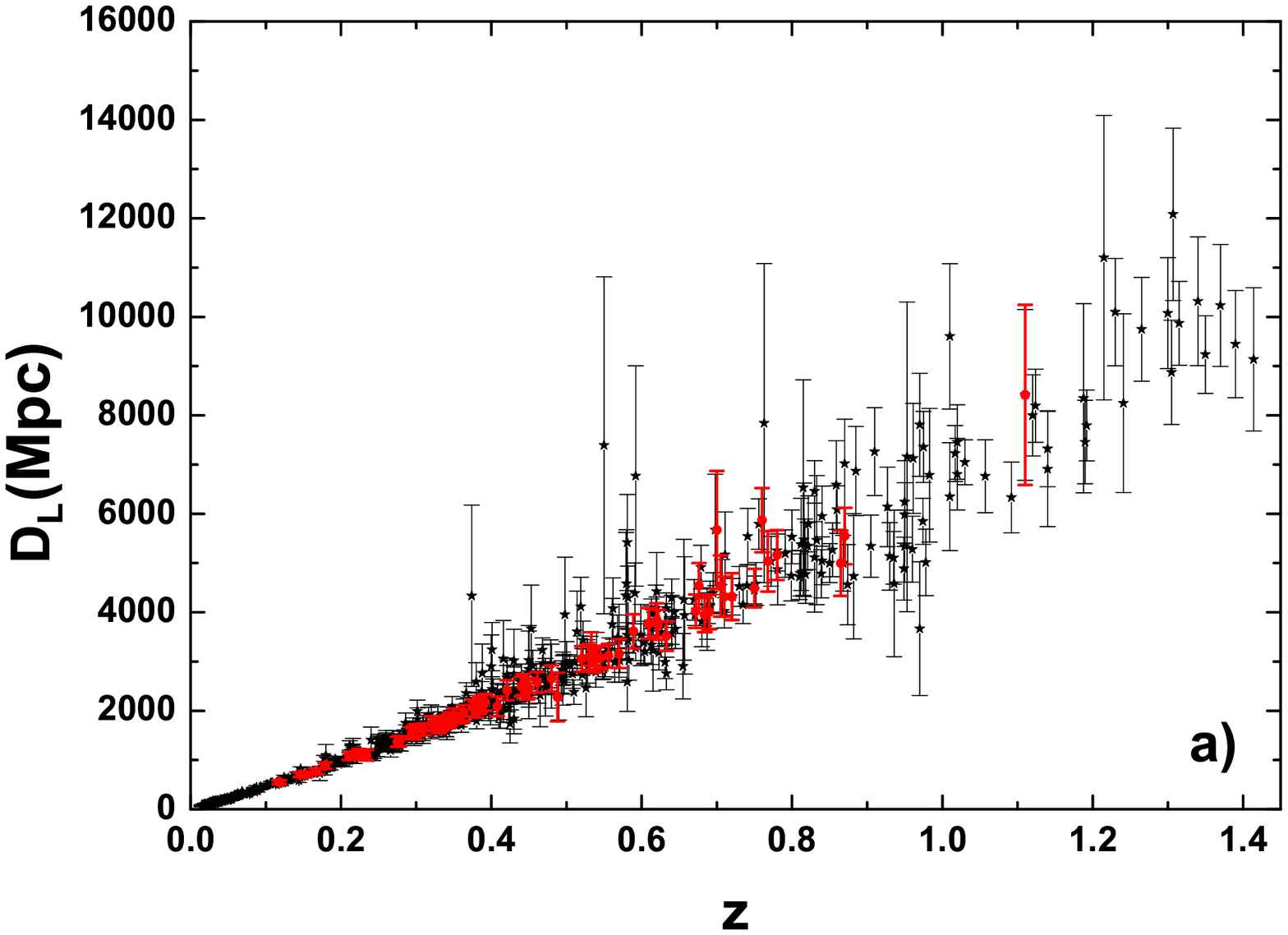}
\hspace{0.3cm}
\includegraphics[width=0.47\textwidth]{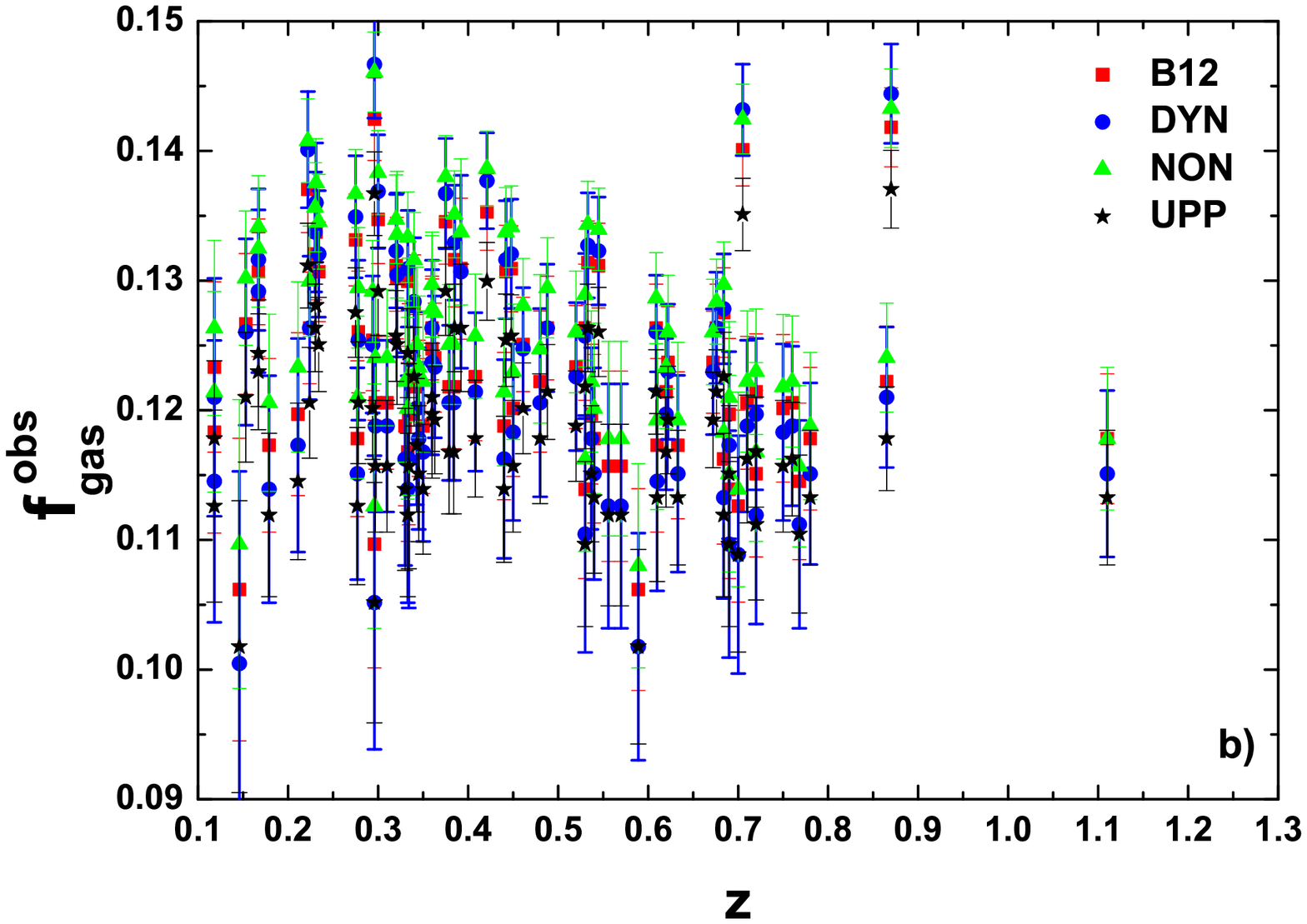}
\caption{In Fig. (1a) we plot the SNe Ia from Union2.1 compilation (black stars) and the points used in our analyses (red circles, see Eq.\ref{eq:dlsigdl}). In Fig. (1b) we plot the 82 gas mass fractions calculated from Ref.\cite{has}. We ruled out 9 galaxy clusters  from original 91 data points due to they do not have SNe Ia pairs with $\Delta z\leq 0.005$.}
\end{figure*}

\section{Samples}

The SZE gas mass fraction data used in this paper were obtained from the cluster mass measurements of the ACT \cite{has} in the redshift range $0.12<z<1.36$ detected via the SZE at 148 GHz. The original sample contains 91  galaxy cluster masses. In order to estimate the galaxy cluster mass, the ACT team adopted a one-parameter family of Universal Pressure Profiles (UPP) as a baseline model for the intracluster gas pressure profile \cite{arn}. The  galaxy cluster masses were measured within a characteristic radius at which the enclosed mean density is 500 times the critical density at the cluster redshift, $M_{500}$. The ACT team also used others three scaling relations to estimate $M_{500}$, which are based on: i) structure formation simulations \cite{bod}, where the density and temperature of the intracluster are modeled as a virialized ideal gas ($M^{B12}_{500}$), ii) a non-thermal pressure and adiabatic model for the gas ($M^{non-thermal}_{500}$) \cite{tra} and iii)  a dynamical estimate of the cluster mass 
using the galaxy velocity dispersions ($M^{dyn}_{500}$) \cite{sif}. Finally, from the total mass it is possible to obtain $f_{gas}$ using the following semi-empirical relation discussed by Ref.\cite{cha}:

\begin{equation}
f^{obs}_{gas} = 0.132 + 0.039 \log M_{15},
\end{equation}
where $M_{15}$ is the cluster total mass $M_{500}$ in units of $10^{15}h^{-1}M_{\odot}$ (see Fig.1b). {  This relation was obtained from dozens of clusters of galaxies in $z<0.09$ (see Table 2 in \cite{cha}) with mass range of $10^{14}-10^{15}h^{-1}M_{\odot}$, which clearly suggested an approximately linear trend of $f^{obs}_{gas}$ with $\log M$. The uncertainties of the coefficients are negligible if compared to uncertainties of the masses in our analyses ($20$-$30\%$). We extrapolate the Eq.(1)  up to $z=1.36$ based on the most recent hydrodynamical simulations that show no significant gas mass fraction evolution with redshift when a $r_{500}$ is used \cite{pla,batt}. It is also important to comment that this relation was obtained from X-ray surface brightness observations, therefore, in this initial approach, we neglect this bias.} 

We also consider a sub-sample of observational measurements of SNe Ia from the original 580 data points of Ref.\cite{suz}, the so-called Union2.1 compilation. The SNe Ia points are  in the redshift range $0.015 < z < 1.43$. The redshifts of SNe Ia were carefully chosen to match the ones of galaxy clusters. In this way, we consider the SNe Ia Union2 compilation \cite{suz} and the galaxy clusters compiled in Ref.\cite{has} as follows: for each galaxy cluster, we select  SNe Ia  with redshifts obeying the  criteria  $|z_{cluster} - z_{SNe}| \leq 0.005$. {  We find 2-6 SNe Ia for each galaxy cluster}. This criteria resulted in 82 galaxy clusters and 82 SNe Ia sub-samples  that  matched this criterion, i. e., 9 galaxy clusters were ruled out from our analyses. This criterion allows us to have some SNe Ia for each galaxy cluster and so we can perform  a  weighted average with them in order to minimize the scatter observed on the Hubble diagram by calculating the following weighted average:

\begin{equation}
\begin{array}{l}
\bar{\mu}=\frac{\sum\left(\mu_{i}/\sigma^2_{\mu_{i}}\right)}{\sum1/\sigma^2_{\mu_{i}}} ,\\
\sigma^2_{\bar{\mu}}=\frac{1}{\sum1/\sigma^2_{\mu_{i}}}.
\end{array}\label{eq:dlsigdl}
\end{equation}.
As is largely known, the distance moduli, $\mu$, of Union2.1 SNe Ia compilation are dependent on the choice of the Hubble parameter  $H_0=70$km/s/Mpc as well as of the $\omega$CDM cosmological model, leaving our results somewhat dependent of a class of cosmological model (see Fig.1a).

\section{Method}

Our method is based on the results from Ref.\cite{hol4}. In that work, the authors showed that the gas mass fraction via SZE observations is dependent on the fine structure constant. In order to clarify the method used, we describe below some fundamental aspects from their results. {  The spherical $\beta$ model is used in this section only for simplicity but without loss of generality for the method proposed.} 

\subsection{Fine structure constant and SZE observations}

The SZE can be expressed for a dimensionless frequency $x \equiv h\nu(z)/k_B T_{CMB}(z)$ as a temperature change $\Delta T(z)$ relative to the CMB temperature $T_{CMB}(z)$ such as:

\begin{equation}
\label{eq:thermal_sz}
\frac{\Delta T(z)}{T_{CMB}(z)}  = f(x,T_e) \int \!\! \, \sigma_T n_e
\frac{k_{\rm B} T_e}{m_e c^2}\,d\ell,
\end{equation}
where $n_e$ and $T_e$ are the electron number density and the gas temperature, respectively, $k_{\rm B}$ the Boltzmann constant, $\sigma_T={8\pi \hbar^2 \alpha^2}/{3 m_e^2 c^2}$ is the Thomson scattering cross-section of the electron, where $\hbar$ is the Planck constant divided by 2$\pi$, $m_e$ is the electronic mass, $c$ is the speed of light and the integral is along the line of sight. The function $f(x,T_e)$ contains the frequency dependence, $\nu$, of the SZE and it can be express as:

\begin{equation}
\label{eq:sze_freq_dependence}
f(x,T_e)=\left( x \frac{e^x + 1}{e^x - 1} -4 \right)
(1+\delta_{\mbox{\tiny SZE}}(x,T_e))
\end{equation}
where $\delta_{\mbox{\tiny SZE}}(x,T_e)$ is a relativistic correction, written in terms of $k_{\rm B}T_e/m_ec^2$ \cite{ito}. As one may see, since $\nu=\nu_0(1+z)$ and $T_{CMB}=T_{0CMB}(1+z)$, the SZE is redshift independent.

By considering the isothermal $\beta$-model, the electron number density is given by
\begin{equation}
n_e({\mathbf{r}}) = n_0 \left ( 1 + \frac{r^2}{r_c^2} \right )^{-3\beta/2},
\label{eq:single_beta}
\end{equation}
where  $r$ is the radius from the center of the cluster, $r_c$ is the core radius of the intracluster medium (ICM) and $\beta$ is a power law index. Under Eq.(3), the SZE decrement profile takes simple analytic forms
\begin{eqnarray}
\Delta T & = & \Delta T_0 \left ( 1 + \frac{\theta^2}{\theta_c^2} \right)
^{(1-3\beta)/2}, \label{eq:easy_sz_signal}
\end{eqnarray}
where  $\Delta T_0$ is the central thermodynamic SZE temperature decrement/increment, and $\theta_c$ is the angular core radius of the cluster.  In this way, the central electron density can be expressed as \cite{gre}:
\begin{equation}
n_0 = \left( \frac{\Delta T_0 \,m_e c^2 \:\Gamma(\frac{3}{2}\beta)}{f_{(x, T_e)}
  T_{CMB} \sigma_T\, k_B T_e D_A \pi^{1/2} \:\Gamma(\frac{3}{2}\beta -
  \frac{1}{2})\, \theta_c} \right)
\label{eq:sz_ne0}
\end{equation}

The mass of gas, inside the radius r, is obtained by integrating the best-fit 3D gas density profile:
\begin{equation}
\Mgas(r) = A \int_{0}^{r/D_A} \left (1+\frac{\theta^2}{\theta_c^2}
\right)^{-3\beta/2}\: \theta^2 d\theta,
\label{eq:mgas_single}
\end{equation}
where $A=4\pi \mu_e n_0 m_p\, D_A^3$, and $\mu_e$, the mean molecular weight of the electrons.

On the other hand, under the  hydrostatic equilibrium assumption, isothermality and Eq. (\ref{eq:single_beta}),   $M_{tot}$ is given by \cite{gre}
\begin{eqnarray}
 M_{tot}(<R)=\frac{3\beta k_BT_G}{\mu Gm_H}\left[\frac{R^3}{(r_{c}^{2}+R^2)}\right],
 \label{Mtot}
\end{eqnarray}
where $T_{G}$ is the temperature of the intracluster medium obtained from X-ray spectrum, $\mu$ and $m_p$ are, respectively, the total mean molecular weight and the proton mass and $G$ is the gravitational constant.

Finally, by considering that the gas mass fraction is defined as \cite{sas}:
\begin{equation}
 f_{gas}=\frac{M_{gas}}{M_{tot}},
 \label{eq3.14}
\end{equation}
where $M_{tot}$ is the total mass  and  $M_{gas}$ is the gas mass obtained by  integrating the gas density model. One may shows, by using the expression for the Thompson scattering cross section, that the current gas mass fraction measurements via SZE depend on $\alpha$ as (see Ref.\cite{hol4} for details): 
\begin{equation}
f^{obs}_{gas} \propto {\alpha^{-2} }\;.
\end{equation}

\subsection{Modified CMB temperature law}

It was shown in Ref.\cite{hee} that modifications of gravity generated by a multiplicative coupling of a scalar field to the electromagnetic Lagrangian lead to a breaking of Einstein equivalence principle  as well as to variations of fundamental constants. As a consequence, we can have $\eta \neq 1$,$\Delta\alpha/\alpha \neq 1$ and $\delta \neq 0$. {  In this framework, the CMB temperature law has to be modified to}

\begin{equation}
T_{CMB}(z)=T_0(1+z)\left[1+0.12\frac{\Delta \alpha}{\alpha}\right].
\label{var}
\end{equation}

In previous papers that used SZE observations to put limits on possible $\alpha$ variation  \cite{hol4,hol5}, the SZE observations were performed in 30 GHz, in this band the effect on the SZE from a variation of $T_{CMB}$ is completely negligible. In the sample considered in the present work, the frequency used to obtain the SZE signal in galaxy clusters  was 148 GHz and the effect from a variation of $T_{CMB}$ on the SZE have to be taking into account \cite{mel}.{  So, following Eq.(12), the term $x$ in Eq.(4) have to be modified to 

\begin{equation}
\psi=h\nu_0/(k_BT_{0CMB}\left[ 1 + 0.12(\varphi(z)-1)\right]),
\end{equation} 
where $z$ in Eq.(13) corresponds to galaxy cluster redshift. Note that if $\Delta \alpha/\alpha=\varphi(z)-1=0$, we have $f(x,T_e)=f(\psi,T_e)$. Therefore, by using the Eqs. (7), (10), (11) and (13), {  if $\varphi(z)\neq 1$},  current gas mass fraction measurements via SZE have to corrected by the factor}
\begin{equation}
(f(x,T_e)/f(\psi,T_e))\varphi(z)^{-2}. 
\end{equation}
In order to calculate this ratio we consider the relativistic corrections from Ref.\cite{ito}, calculated up to the fifth order of $kT_e/m_ec^2$. 

The temperature of the galaxy clusters were estimated by using the scaling relation from Ref.\cite{arn}, obtained via ten relaxed galaxy clusters with $z \leq 0.15$, such as $h(z) M=A\left[ kT_e/5keV\right]^\tau$, where $A$ and $\tau$ are, respectively: $4.10 \pm 0.19$ and $1.49 \pm 0.15$. The $h(z)$ parameter corrects the evolution expected in the standard self-similar model. This parameter is  between  $1.05$ and  $1.28$ for the Chandra clusters located at higher redshifts ($0.1 < z < 0.46$). We consider a medium value $\approx 1.20$.

\begin{figure*}
\centering
\includegraphics[width=0.47\textwidth]{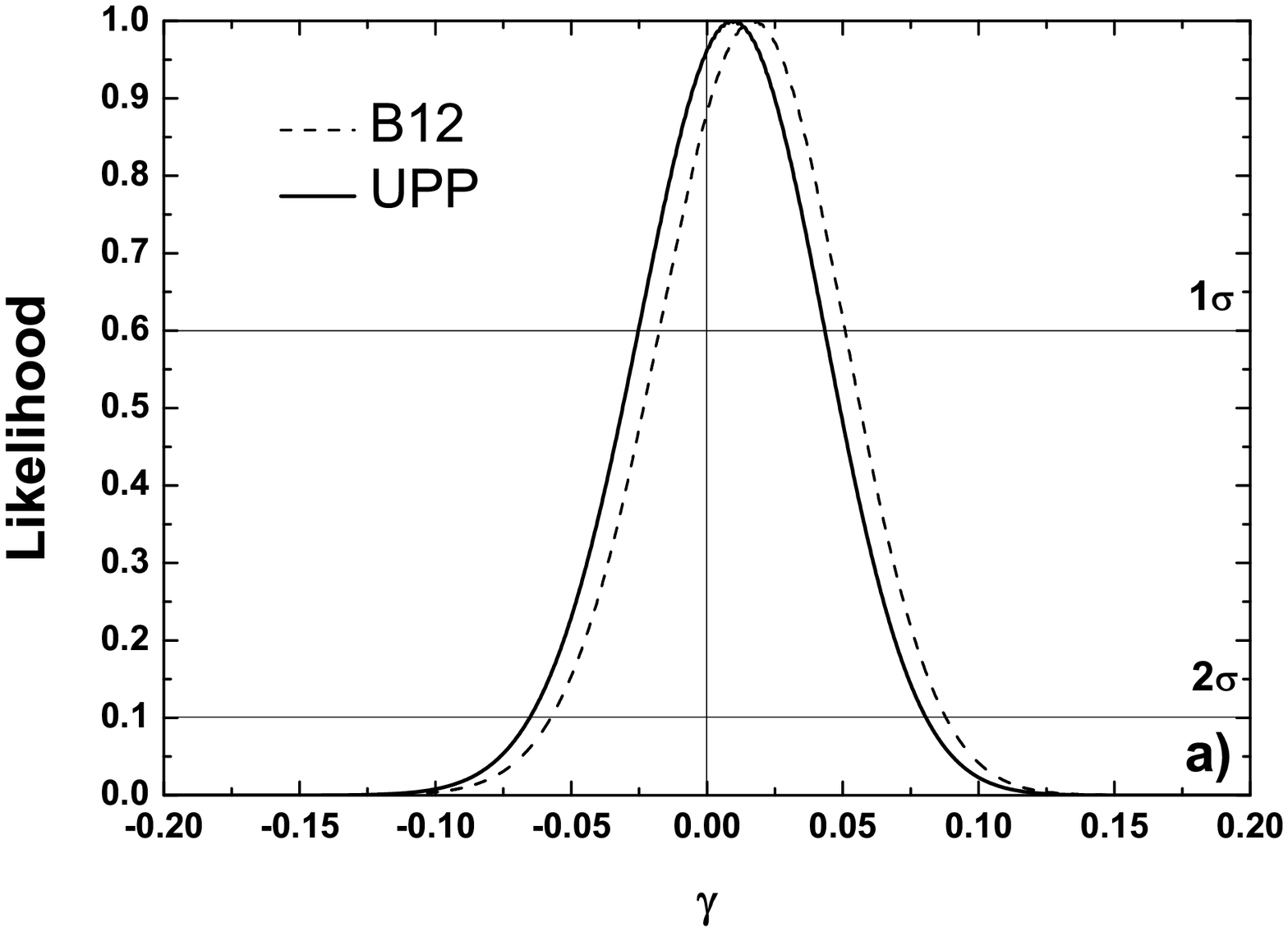}
\hspace{0.3cm}
\includegraphics[width=0.47\textwidth]{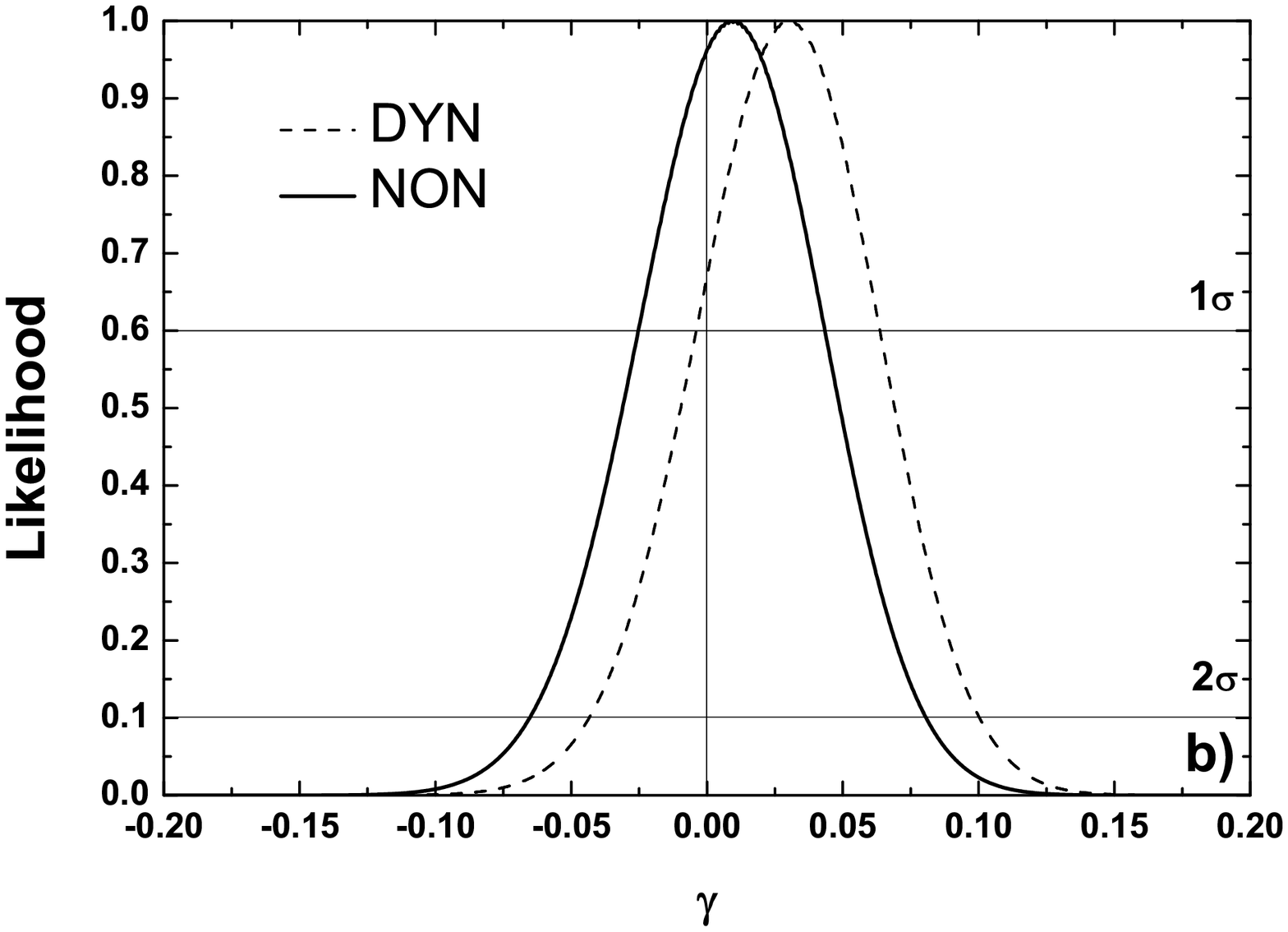}
\caption{ The likelihood functions from our analyses. In Fig.(2a) we plot the results from UPP (solid line) and B12 (dashed line) models. In Fig.(2b) we plot the results from NON (solid line) and DYN (dashed line) models.  
}
\end{figure*}

\subsection{Observational equation for $\varphi(z)$}

The expression of the SZE gas mass fraction used as cosmological tool is \cite{lar}:

\begin{equation}
f^{obs}_{gas}=N\left(\frac{D_A^{*}}{D_A}\right),
\end{equation}
where the symbol * denotes quantities that were obtained by using a fiducial model  in the observations and the parameter $N$ defines the astrophysical modeling of the cluster. Following Eq.(15), this relation must be corrected to:
\begin{equation}
\zeta f^{obs}_{gas}\varphi(z)^{-2}=N\left(\frac{D_A^{*}}{D_A}\right),
\end{equation}
where $\zeta = (f(x,T_e)/f(\psi,T_e))$.

As we aim to put limits on $\varphi(z)$, we could consider the validity of the CDDR - so $D_L(1+z)^{-2}/D_A  =\eta=1$ - and use distance moduli from a Union2.1 SNe Ia compilation to obtain $D_L$, leading to bounds on $\varphi(z)$. However, from Eq.(12),  the CDDR has to be modified to $D_L(1+z)^{-2}/D_A  =\varphi(z)^{1/2}$ before use it.

After performing simple algebraic operations one obtains:
\begin{equation}
\varphi^{obs}(z)=\left(\frac{\zeta f^{obs}_{gas}10^{\frac{\bar{\mu}-25}{5}}}{ND^*_L}\right)^{2/5},
\end{equation}
where we use  $D_L(z)=10^{\frac{\bar{\mu}-25}{5}}$ Mpc.

In our analyses, we focus on the dilaton runaway models (see more details in Refs.\cite{dam}) where the relevant parameter for studying the variation of $\alpha$  is the coupling of the dilaton field to hadronic matter.{  We are interested in the evolution of the dilaton and a reasonable approximation in the redshift range used in our analyses is to linearize the field evolution, such as
\begin{equation}\label{evolslow}
\frac{\Delta\alpha}{\alpha}(z)\approx\, -\frac{1}{40}\beta_{had,0} {\phi_0'}\ln{(1+z)}=-\gamma\ln(1+z)\;,
\end{equation}
or, equivalently, $\varphi(z)=1-\gamma\ln(1+z)$, where $\gamma = \frac{1}{40}\beta_{had,0} {\phi_0'}$ with $\phi_0'= \frac{\partial \phi}{\partial \ln a}$ being the scalar field at present time and $\beta_{had,0}$ being the current value of the coupling between the dilaton and hadronic matter. }

\section{Analyses and results} 

We evaluate our statistical analyses by defining the likelihood distribution function ${\cal{L}} \propto e^{-\chi^{2}/2}$, where
\begin{equation}
\label{chi2} 
\chi^{2} = \sum_{i = 1}^{82}\frac{{\left[(1-\gamma\ln{(1+z)}) - \varphi_{i, obs} \right] }^{2}}{\sigma^{2}_{i, obs}},
\end{equation}
with {$\varphi^{obs}(z)=\left(\frac{\zeta f^{obs}_{gas}10^{\frac{\bar{\mu}-25}{5}}}{ND^*_L}\right)^{2/5}$ and $\sigma^2_{i,obs}$ is  the uncertainty associated to observational quantities: $f^{obs}_{gas}$, $\bar{\mu}$ and $kT_e$. {  The parameter $N$ carries all the information
about the matter content in the cluster, such as stellar mass fraction, non-thermal pressure and the depletion parameter, which indicates the amount of cosmic baryons that are thermalized within the cluster potential \cite{all}. From hydrodynamical simulations this quantity does not have significant dependence on redshift. Moreover, since the most of cluster masses used in our analyses are of the same order, $10^{14}M_{\odot}$, we take it as a nuisance parameter so that we marginalize over it}. Following  Ref.\cite{suz} we added  a 0.15 systematic error to SNe Ia data. Constraints on the quantity $\gamma = \frac{1}{40}\beta_{had,0} {\phi_0'}$ are shown in Figs. (2a) and (2b). 

From Fig. (2a) we obtain $\gamma = 0.008\pm 0.035$ and $\gamma = 0.018\pm 0.032$ (at 68.3\% c.l.) for UPP and B12 models, respectively. From Fig. (2b) we obtain $\gamma = 0.01\pm 0.033$ and $\gamma = 0.030\pm 0.033$ (at 68.3\% c.l.) for NON and DYN models, respectively. As one may see, all results are fully compatible each other and with $\phi(z)=1$ or, equivalently, with no variation of fine structure constant $\alpha$.

It is interesting to compare our bounds on $\gamma $ with the limits obtained recently from galaxy clusters and SNe Ia by Refs.\cite{hol4,hol5}. In Ref.\cite{hol4} the authors showed that observations of the gas mass fraction via SZE and X-ray surface brightness of the same galaxy cluster are related by $f_{SZE}=\varphi(z)f_{X-ray}$, where $\varphi(z)=\frac{\alpha}{\alpha_0}$. Using 29 $f_{gas}$ measurements  they found $\gamma = 0.065\pm 0.095$ at 68.3\% (C.L.), in full agreement with our results. In Ref.\cite{hol5} the authors showed that measurements of the SZE combined with observations of the X-ray surface brightness of galaxy clusters for estimating the ADD of galaxy clusters depends on the fine structure constant besides $\eta$ \cite{uza}. By using 25 ADD and current type Ia supernovae observations they found: $\gamma = -0.037\pm 0.0157$ (at 68.3\% c.l.). In this way, no significant indication of variation of $\alpha$ with the present data was found.

\section{Conclusions}

Nowadays, one of the most important fields of research in Cosmology is to investigate the physical assumptions implicit in the cosmological models. In this paper we analyzed a possible variation of the fine-structure constant ($\varphi(z)=\frac{\alpha}{\alpha_0}$), using observations of masses from galaxy clusters, for a special class of runaway dilaton models. The measurements were obtained via the Sunyaev-Zeldovich effect and the masses were obtained for four different scaling relations, named $M^{UPP}_{500}$, $M^{B12}_{500}$, $M^{non-thermal}_{500}$ and $M^{dyn}_{500}$. The gas mass fraction data were then obtained using a semi-empirical relation and combined with SNe Ia measuments.

In order to perform our analysis, we use a data set of 82 points of galaxy cluster mass and SNe Ia measurements, by assuming a limit of $|z_{cluster} - z_{SNe}| \leq 0.005$. The gas mass fraction and the distance moduli from SNe Ia were combined using the CDDR assuming a possible variation of the fine structure constant, thus $D_L(1+z)^{-2}/D_A  =\varphi(z)^{1/2}$. Assuming the evolution of the dilaton as $\frac{\Delta\alpha}{\alpha}(z)=-\gamma\ln(1+z)$, we performed  statistical chi-square analyses in order to obtain the best fit value of $\gamma$ for the different estimates of the galaxy clusters masses.

We found results in complete agreement between each other, with $\gamma^{UPP} = 0.008\pm 0.035$, $\gamma^{B12} = 0.018\pm 0.032$, $\gamma^{NON} = 0.01\pm 0.033$ and $\gamma^{DYN} = 0.030\pm 0.033$, with 68.3\% of confidence level. By comparing our results with others in the literature, we  found that the combination of gas mass fraction via SZE and SNe Ia measurements can produce results as robust as others using different methods.

\section*{Acknowledgments}
RFLH acknowledges financial support from INCT-A and CNPq (No. 478524/2013-7, 303734/2014-0). JSA is supported by CNPq, FAPERJ and INEspa\c{c}o. L. R. Cola\c{c}o is supported by CAPES.

\label{lastpage}
\end{document}